# Sobre la falsa elección entre la regulación y la innovación. Ideas para la regulación mediante un uso responsable de la inteligencia artificial en investigación y educación.

[**On the false election between regulation and innovation. Ideas for regulation through the responsible use of artificial intelligence in research and education**.]

AI Hub-CSIC / EduCaixa, Escuela de Verano, Auditorio CaixaForum, Zaragoza, 4 de julio de 2025

Pompeu Casanovas

Profesor de Investigación en el Instituto de Investigación en Inteligencia Artificial (IIIA-CSIC), Barcelona

Adjunct Professor, La Trobe University Law School, Melbourne

Investigador vinculado al Institut de Dret i Tecnologia de la Universitat Autònoma de Barcelona (IDT-UAB)

pompeu.casanovas@iiia.csic.es

**Resumen.** Este breve ensayo es una reelaboración de las respuestas ofrecidas por el autor en la Sesión de Debate de la Escuela de Verano de AIHUB (CSIC) y EduCaixa el 4 de julio de 2025, organizada por Marta García-Matos y Lissette Lemus y coordinada por Albert Sabater (OEIAC, UG), con la participación de Vanina Martínez-Posse (IIIA-CSIC), Eulàlia Soler (Eurecat) y Pompeu Casanovas (IIIA-CSIC). Albert Sabater planteó tres preguntas: (1) ¿Cómo pueden los marcos regulatorios priorizar la protección de derechos fundamentales (privacidad, no discriminación, autonomía, etc.) en el desarrollo de IA, sin caer en la falsa dicotomía entre regulación e innovación? (2) Ante los riesgos de la IA (sesgos, vigilancia masiva, manipulación), ¿qué ejemplos de regulaciones o políticas han demostrado que es posible fomentar innovación responsable, anteponiendo el interés público a la rentabilidad, sin ceder a la presión competitiva de actores como China o EE. UU? (3) En un escenario donde los EE. UU priorizan la flexibilidad, ¿qué mecanismos podrían garantizar que la cooperación internacional en IA no se convierta en una carrera hacia el abismo en derechos, sino en un estándar global de responsabilidad? El artículo trata de responder a estas tres preguntas, y concluye con algunas reflexiones sobre la relevancia de las respuestas para la educación y la investigación.

**Abstract**. This short essay is a reworking of the answers offered by the author at the Debate Session of the AIHUB (CSIC) and EduCaixa Summer School, organized by Marta García-Matos and Lissette Lemus, and coordinated by Albert Sabater (OEIAC, UG), with the participation of Vanina Martínez-Posse (IIIA-CSIC), Eulàlia Soler (Eurecat) and Pompeu Casanovas (IIIA-CSIC) on July 4[th] 2025. Albert Sabater posed three questions: (1) How can regulatory frameworks prioritise the protection of fundamental rights (privacy, non-discrimination, autonomy, etc.) in the development of AI, without falling into the false dichotomy between regulation and innovation? (2) Given the risks of AI (bias, mass surveillance, manipulation), what examples of regulations or policies have demonstrated that it is possible to foster responsible innovation, putting the public interest before profitability, without giving in to competitive pressure from actors such as China or the US? (3) In a scenario where the US prioritizes flexibility, what mechanisms could ensure that international cooperation in AI does not become a race to the bottom in rights, but rather a global standard of accountability? The article attempts to answer these three questions and concludes with some reflections on the relevance of the answers for education and research.

# 1. Introducción

Buenos días a todos. Albert Sabater nos ha planteado tres cuestiones sobre regulación e innovación. Intentaré ofrecer en esta breve intervención algunas ideas para el debate.[1]

Vamos a dejar de lado hoy la "disrupción" o alteración que se ha producido en el mundo jurídico por el hecho de la utilización de IA en la producción y circulación de bienes. Cada semana se están interponiendo nuevas demandas sobre el uso de textos, imágenes o sonidos protegidos por propiedad intelectual para el entrenamiento de sistemas de LLM sin permiso del autor. Por ejemplo, el 23 de junio de 2025 un juez federal de California dictaminó en *Bartz et al v. Anthropic* la legalidad del uso de los libros adquiridos por compraventa y la ilegalidad de los que no lo habían sido. El 25 de junio, otro juez federal dictaminó el uso legítimo (*fair use*) de textos en función de la propiedad en *Kadrey v. Meta Platforms* basándose en la *Copyright Act*.

Las demandas empezaron pronto, inmediatamente después del lanzamiento público de ChatGPT el 30 de noviembre de 2022. El 27 de diciembre de 2023 fue el NY Times quien emprendió acciones legales contra OpenAI y Microsoft.[2] Había razones para ello y, de hecho, las resoluciones judiciales probablemente obliguen a replantear técnicamente la estrategia interna de OpenAI.[3] Este caso acaba de empezar y muy probablemente tendrá implicaciones importantes para la propiedad y gestión de datos.[4] Existe ya una base de datos de casos judiciales en los que la IA está involucrada, *DAIL-the Dababase of AI Litigation*.[5]

Los tribunales europeos también han empezado a buscar soluciones, insistiendo en el aspecto asistencial y no decisor de la IA. Según la sentencia del Tribunal de Justicia de la Unión Europea (TJUE) del 7 de diciembre de 2023 que abordaba la cuestión planteada por el Tribunal Administrativo de Wiesbaden sobre las actuaciones de SCHUFA, una empresa de *scoring*, está prohibido efectuar decisiones bancarias o de concesión de crédito basadas en las predicciones de patrones de comportamiento futuro calculadas por sistemas generativos. Éste es también el espíritu del Reglamento Europeo de Protección de

---

[1] Vid. el registro y edición en vídeo de la sesión en "Sobre la falsa elección entre la regulación y la innovación", Sesión de Debate de la Escuela de Verano de AIHUB (CSIC) y EduCaixa el 4 de julio de 2025.

[2] Se trata de una demanda interpuesta por *The New York Times*, *The New York Daily News* y ocho periódicos más, junto con el *Center for Investigative Research Inc.,* contra OpenAI, entidades relacionadas y Microsoft Corp. como socio inversor. *The New York Times Company, et al. v. Microsoft Corporation, et al,* S.D.N.Y. Case No. 1:23-CV-11195.

[3] Los modelos GPT y Claude utilizaban entrenamiento de rechazo y filtros de salida para evitar la salida textual (*verbatim output*) de los artículos memorizados. Cf. Joshua Freeman, Chloe Rippe, Edoardo Debenedetti, Maksym Andriushchenko (2024). "Exploring Memorization and Copyright Violation in Frontier LLMs: A Study of the New York Times v. OpenAI 2023 Lawsuit", https://arxiv.org/abs/2412.06370 . Según los autores, los modelos OpenAI son actualmente menos propensos a la elicitación de memorización que los modelos Meta, Mistral y Anthropic.

[4] Cf. Jeffrey M. Kelly, Scott N. Sherman, Adrianne Cleven, Matt Gorga [de Nelson Mullin (10 de julio 2025), "From Copyright Case to AI Data Crisis: How The New York Times v. OpenAI Reshapes Companies' Data Governance and eDiscovery Strategy". Como subrayan los autores, la orden judicial emitida el 13 de mayo de 2025 por la magistrada Ona T. Wang exigiendo que OpenAI conserve todos los registros de conversaciones de ChatGPT afecta a más de 400 millones de usuarios en todo el mundo y sus implicaciones van mucho más allá del litigio, puesto que "los líderes empresariales que implementan soluciones de IA se enfrentan ahora a preguntas sin precedentes sobre la soberanía de los datos, la preparación de juicios y el equilibrio entre la innovación y la observación del derecho".

[5] Cf. DAIL – the Database of AI Litigation

Datos (2016)[6] y de la reciente Ley de la IA (2024). Había sido anticipada por algunos escándalos notorios en las predicciones en materia criminal en EEUU, como las de COMPAS y programas afines.[7] Además, como vimos en el Seminario de Retos AIHUB-EduCaixa de Camilo Chacón y Martín Sánchez-Fibla (3 de julio 2025), si es cierto que la IA generativa es capaz de inventar nuevos algoritmos, es decir, propiamente de *crear*, puede plantear problemas a lo que el derecho de patentes considera una invención. El caso DABUS, planteado por Stephen Thaler sobre la posible personalidad jurídica atribuible a la IA, aún se está debatiendo en los tribunales de varios continentes.[8] Aunque, de momento, con poco recorrido.[9]

Pero hoy no vamos a centrarnos en estos aspectos más cercanos a la visión clásica del derecho, si podemos decirlo así. Albert Sabater nos ha propuesto un debate más amplio sobre la regulación de la IA partiendo de la tesis de "la falsa elección entre la regulación digital y la innovación", un planteamiento avanzado por Anu Bradford[10], experta norteamericana en derecho internacional y autora the *The Brussels Effect* (2020) y *Digital Empires: The Global Battle to Regulate Technology* (2024).

En uno de sus últimos artículos en *Foreign Affairs*, Bradford recuerda que "incluso los poderosos gigantes tecnológicos como Apple, Google, Meta y Microsoft usan el reglamento europeo de protección de datos como política de privacidad global"[11] y, para hacer frente a la política norteamericana de expansión tecnológica, recomienda proseguir la construcción de un mercado digital que pueda situar los derechos humanos y cívicos en primer lugar al mismo tiempo que incentiva la innovación de las empresas tecnológicas. Es decir, propone no ceder ante las presiones de Trump, fortaleciendo la dimensión obligatoria del estado de derecho: "Europa no debe doblegarse ante la presión de Estados

---

[6] El art. 22, apartado 1, del Reglamento General de Protección de Datos (RGPD) establece que cualquier persona tiene el derecho de no ser sujeta a una decisión basada únicamente en el tratamiento automatizado, incluida la elaboración de perfiles, que tenga efectos jurídicos sobre ella o la afecte de manera similar. Cf. *Reglamento (UE) 2016/679 del Parlamento Europeo y del Consejo de 27 de abril de 2016 relativo a la protección de las personas físicas en lo que respecta al tratamiento de datos personales y a la libre circulación de estos datos y por el que se deroga la Directiva 95/46/CE* (Reglamento general de protección de datos) (Texto pertinente a efectos del EEE)

[7] Cf. Sascha van Schendel (2019). "The challenges of risk profiling used by law enforcement: Examining the cases of COMPAS and SyRI". Leonie Reins (ed.), *Regulating new technologies in uncertain times*. Cham: Springer, pp. 225-240.

[8] DABUS (*Device for the Autonomous Bootstrapping of Unified Sentience*) planteó en diversas oficinas de patentes en quince jurisdicciones (incluyendo USA, UK, Australia, Alemania, Sudáfrica y Japón) que los sistemas de IA podían crear innovaciones patentables y que si era así son éstos y no los humanos los que deberían poder figurar como autores. Thaler alegó que DABUS había producido dos inventos sin su intervención: (i) un contenedor de alimentos y bebidas que "permite sujetar firmemente el envase con un brazo robótico"; (ii) y una luz que parpadea de una manera única para atraer la atención del público durante situaciones de emergencia. Cf. A. Saravanan and D. Prasad (2025). "AI as an Inventor Debate under the Patent Law: A Post-DABUS Comparative Analysis", *European Intellectual Property Review*, 47 (1): 26-39 (forthcoming), SSRN: https://ssrn.com/abstract=5053108 o http://dx.doi.org/10.2139/ssrn.5053108.

[9] Vid. un anàlisis de la aplicación de conceptos jurídicos a los sistemas de IA en Pompeu Casanovas y Pablo Noriega, (2025). "Governance of Artificial Agency and AI Value Chains: A Few Remarks on Autonomy from a Legal and Ethical Approach", in López Castro, M. Cebral, P. Jiménez-Schlegl, *Regulating Autonomy: Ethics, Values and Governance in Intelligent Hybrid Systems*, Cham: Springer (forthcoming).

[10] Anu Bradford (2024). "The false choice between digital regulation and innovation." *Northwestern University Law Review* 119 (2024): 377-452.

[11] Anu Bradford, R. Daniel Kelemen, and Tommaso Pavone "Europe Could Lose What Makes It Great U.S. Pressure and Domestic Rancor Threaten the EU's Regulatory Superpower", *Foreign Affairs*, April 21, 2025

Unidos y el rencor doméstico (*domestic rancor*)". Es la dimensión vertical (coactiva) del estado de derecho (*rule of law*) expresado en el diagrama (Fig.1).

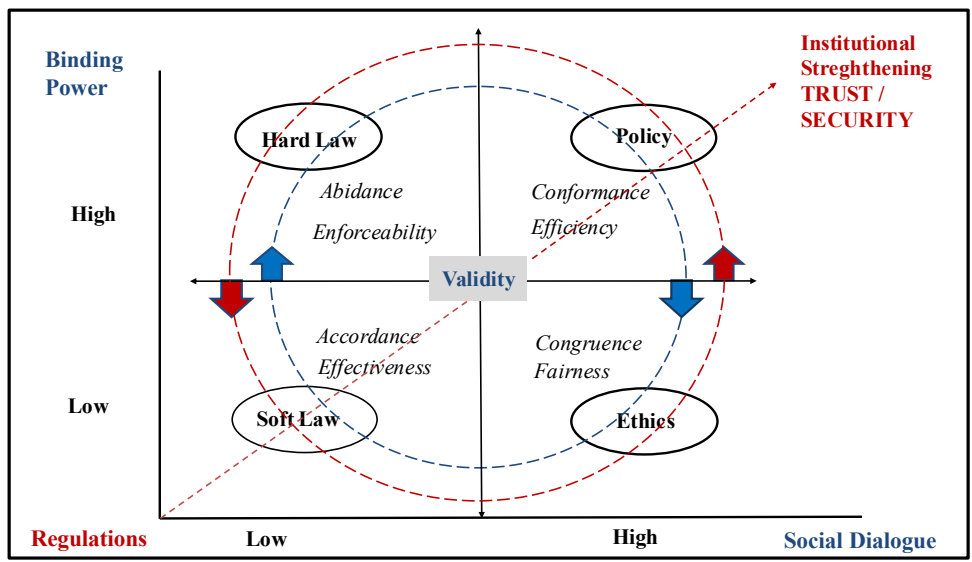

Fig. 1. Esquema de la regulación en el estado de derecho (*rule of law*). Fuente: Casanovas, Hashmi y de Koker (2021).

El esquema comprende un eje vertical de poder ejecutivo y otro horizontal, basado en el diálogo y la negociación.[12] Ambos deben tenerse en consideración en la construcción de modelos de regulación y en la aplicación de la ley. La proporcionalidad entre ambos ejes ha sido objeto de debate entre políticas reguladoras y desreguladoras. Esto tampoco es nuevo. En los años treinta se produjo un debate acalorado en Estados Unidos acerca de la '*regimentation*' producida por el programa legislativo del New Deal. Los que se oponían a él (y a los iusrealistas que lo apoyaban) lo utilizaron como arma.[13] En inglés norteamericano 'to regiment' tiene una connotación coercitiva muy fuerte. Equiparar 'regular' a 'regimentar' (o 'reglamentar') es una de las estrategias políticas de signo conservador para oponerse a cualquier iniciativa normativa, sobre todo estatal. Ahora hay que considerar tanto que los agentes y actores (*stakeholders*) no son solo humanos, como muestra el carácter informacional de las transacciones operadas sobre datos y metadatos. El objeto y ámbito de regulación del estado de derecho, pues, se dirige a una sociedad *híbrida*, que no es exactamente la misma que teníamos hace sólo treinta años.

Bradford es conocida por su tripartición del marco general de regulación entre (i) el estado (la ultraregulación política china), (ii) el mercado (la desregulación norteamericana) y (iii) los derechos (la política europea de extensión de las limitaciones de la privacidad a las grandes corporaciones tecnológicas norteamericanas, especialmente desde 2012 a través del GDPR). Este último es el marco que Europa debería sustentar para conservar su "superpoder regulatorio", en expresión de Bradford.[14] Según este planteamiento, la forma de hacerlo es reforzar su poder de ejecución (*enforcement*) en lugar de ceder a la tentación del relajamiento o tolerancia regulatoria (*forbearance*), una política anterior a la crisis del 2008 impulsada bajo la presidencia de en la Comisión de José Manuel Barroso y vuelta a reactivar por el *Informe Draghi*[15] de septiembre de 2024 (también por los *Informes Letta*[16] y *Eurostack*[17]). Bradford insiste en la necesidad de reforzar los poderes de ejecución, especialmente porque las corporaciones tecnológicas no pueden permitirse perder el mercado europeo y por la tendencia de otros países, como Brasil, de reforzar el marco de regulación en la dirección señalada por la UE.[18]

---

[12] Cf. Casanovas, Pompeu, Mustafa Hashmi, and Louis de Koker (2021). "The rule of law and compliance: Legal quadrant and conceptual clustering." In *International Workshop on AI Approaches to the Complexity of Legal Systems*, pp. 215-229. LNCS 13048. Cham: Springer International Publishing, 2021.
[13] Cf. Henry S. Commager (1934). "Regimentation: A New Bogy." *Current History (1916-1940)* 40 (4): 385-391.

El problema que nos ocupa en la discusión es cómo, de qué manera esto puede ser posible. El marco general sugerido por Bradford ya ha recibido las críticas de algunos juristas expertos en privacidad, quienes han puesto de manifiesto que el optimismo regulador de la Unión Europea puede provocar un efecto contrario al que buscaba. No siempre el afán regulatorio produce efectos positivos.[19] Y me temo que, sólo dos años después, hallar una respuesta es aún más difícil puesto que los presupuestos sobre la paz, el mercado global y el desarrollo, impulsados por la idea de globalización de los años noventa han cedido a un panorama mucho más sombrío, de política real bismarkiana, dónde la fuerza militar organizada prevalece sobre los ideales de la sociedad civil (incluyendo la justicia en los intercambios y en el mercado).

Hay que poner de manifiesto, además, que la reacción de la Unión Europea a la imposición de aranceles por parte de Trump no ha fortalecido su función de reguladora internacional. Al contrario, ha puesto de manifiesto una debilidad fundamental de la propia Unión que le ha hecho perder posiciones a nivel geoestratégico global. En un nuevo orden mundial, no está para dar lecciones.

Pero vayamos ya a las cuestiones planteadas por Albert Sabater.

---

[14] "And behind this power also lies Europe's normative appeal: unlike the Chinese model of statist regulation or the traditional American model of deregulatory market capitalism, the EU's regulatory approach puts the rights of citizens first."

[15] Mario Draghi (2024). The Draghi report: A competitiveness strategy for Europe (Part A); The Draghi report: In-depth analysis and recommendations (Part B). EU Commission. Véase también, más adelante, las notas 60 y 61 sobre la reflexión de Draghi de 16 de septiembre de 2025 sobre su propio texto, un año después.

[16] Enrico Letta (2024). Much more than a market. Speed, Security, Solidarity. Empowering the Single Market to deliver a sustainable future and prosperity for all EU Citizens. EU Commission, April 2024.

[17] Francesca Bria, Paul Timmers, Fausto Gernone (2025). EuroStack – A European Alternative for Digital Sovereignty, Gütersloh: Bertelsmann Stiftung, February 2025.

[18] Vid. el diálogo entre Anu Bradford, Kate Klonick, Kevin Frazier "Scaling Laws: Contrasting and Conflicting Efforts to Regulate Big Tech: EU v. U.S.", *Lawfare*, 4 de septiembre 2025.

[19] Vagelis Papakonstantinou, Paul De Hert (2022). "The Regulation of Digital Technologies in the EU: The Law-Making Phenomena of 'Act-ification','GDPR Mimesis' and 'EU Law Brutality'." *Technology and Regulation Journal* (2022); Pagallo, Ugo, "Why the AI Act Won't Trigger a Brussels Effect" (December 16, 2023). *AI Approaches to the Complexity of Legal Systems*, Forthcoming, Available at SSRN: https://ssrn.com/abstract=4696148 or http://dx.doi.org/10.2139/ssrn.4696148.

## 2. Primera pregunta

**¿Cómo pueden los marcos regulatorios priorizar la protección de derechos fundamentales (privacidad, no discriminación, autonomía, etc.) en el desarrollo de IA, sin caer en la falsa dicotomía entre regulación e innovación?**

Voy a proponer algunas sugerencias que se resumen en una: hay que cambiar la idea de qué significa regular y la forma cómo se construyen los modelos de regulación (incluyendo el sistema normativo de los derechos). Es decir, *tenemos que aceptar el reto de la reconstrucción del espacio público en y desde el ámbito de la tecnología y la IA. Y este cambio debería hacerse desde la base y desde el conocimiento orientado a la creación de cadenas de valor no solamente económicas sino sociales en las transacciones y los procesos de información*. En concreto, deberían aprovecharse las infraestructuras creadas por los espacios comunes europeos de datos (agricultura, patrimonio cultural, energía, finanzas, pacto verde, salud, industria, idiomas, medios de comunicación, movilidad, administración pública, investigación e innovación, habilidades y turismo). Y deberíamos dotarnos de métricas para medir su funcionamiento y avances.

Pero, como ya he dicho, esto no es sencillo. Hay algunas precondiciones que son previas a la construcción de requerimientos (condiciones) para la construcción de modelos. En suma:

(i) controlar el comportamiento de las élites (i.e. el comportamiento e ideología de los CEOs de las grandes corporaciones tecnológicas);

(ii) definir claramente el ámbito, modelo e implementación del sistema normativo;

(iii) tomar en serio la posibilidad de regular esta sociedad híbrida (entre humanos y máquinas) mediante mecanismos semiautomáticos de Cumplimiento *mediante* (*through*) Diseño y no Cunplimiento *por* (*by*) Diseño (CtD vs. CbD) desde una perspectiva de gobernanza jurídica (*legal governance*) y no de mera "aplicación del derecho";

(iv) construir ecosistemas éticos y jurídicos que puedan implementarse en tiempo real, en lugar de seguir la secuencia clásica del siglo XIX sobre la redacción, interpretación y aplicación de leyes);

(v) volver a situar la ética en el centro de los modelos regulatorios (al margen de jurisdicciones particulares);

(vi) confiar en la reacción y proactividad de los ciudadanos en la aceptación, cumplimiento y mantenimiento de los ecosistemas;

(vii) reconocer y gestionar los conflictos que subyacen en la geopolítica estadounidense, rusa y china (i.e. las tensiones por el poder y el control de la población deben situarse donde están y no ser negadas). Los conflictos de interés, económicos y políticos están en el centro de la actuación internacional de los estados, y me temo que esto seguirá siendo así.

Y, en relación al campo educativo, podríamos añadir aún:[20]

(viii) ampliar el marco educativo digital de IA, para planificar espacios educativos de alfabetización jurídica y ética en y con la IA, que permita un uso responsable y una colaboración eficaz y apropiada en todos los ámbitos.

(ix) estimular situaciones dinámicas y reales de participación social con IA, que reformulen su papel y función en la sociedad, más allá de la creación de canales comunicativos institucionales.

Más aún que la crisis de 2008, hay un punto de incidencia transcendental en el cambio de política de las élites, las corporaciones y el estado en la disminución y restricción del espacio público: *el pánico y las oportunidades de negocio en los conflictos bélicos que se originaron a partir del 11 de septiembre en 2001*. Este punto de incidencia, además, ha sido absorbido por el espacio social. Tres ejemplos:

(i) El giro hacia las empresas privadas de armamento y ciberseguridad en detrimento de las públicas en las políticas de defensa de la administración norteamericana; por ejemplo el escándalo *Trailblazer Project* vs. *Thinthread* en 2002 denunciado internamente por William Binney, J. Kirke Wiebe, Edward Loomis y por Diane Roarke (miembro del *House Permanent Select Committee on Intelligence* ).[21]

(ii) El cambio en la actitud de los CEOS en la gestión y dirección de las corporaciones tecnológicas. Ahora ya han salido a la luz las "creencias de lujo" (para decirlo con Henderson) de Peter Thiel, CEO de PayPal, y de Alexander Karp, CEO de Palantir, ambos formados en la raíz de la filosofía europea francesa (René Girard) y alemana (Theodor Adorno y Jürgen Habermas)[22]; y ambos defensores de la agresividad (Karp) y de la mímesis —incluyendo la violencia y la necesidad de chivos expiatorios (Thiel)— como forma de integración social en una nueva versión de la contra-Ilustración o contra-Escuela de Frankfurt.[23]

No hay que equivocarse aquí: existe no solamente una visión expansionista de mercado, sino el diseño de un modelo de sociedad basado en el conocimiento, la ley del más fuerte en innovación y la eliminación de aquellos que lo impidan, sea por su posición como competidores, de enemigos o sencillamente de lastre. Los pobres y los considerados inútiles también entran en esta descripción. Se trata del posicionamiento de una oligarquía del conocimiento.

---

[20] Agradezco a Neus Lorenzo el añadido de estos dos puntos.

[21] Este caso es de dominio público. Cf. Wikipedia, list of whistleblowers: "NSA officials initially joined House Permanent Select Committee on Intelligence staffer Diane Roark in asking the U.S. Department of Defense inspector general to investigate wasteful spending on the Trailblazer Project and the NSA officials eventually went public when they were ignored and retaliated upon. They claim that Thinthread was more focused and thus more effective and lower cost than Trailblazer and subsequent programs, which automatically collected trillions of domestic communications of Americans in deliberate violation of the U.S. Constitution."

[22] Alexander C. Karp. Tesis en Filosofía: *La agresión en el-mundo-de-la-vida: La extensión del concepto de agresión de Parsons mediante la descripción de la conexión entre jerga, agresión y cultura* [*Aggression in der Lebenswelt: Die Erweiterung des Parsonsschen Konzepts der Aggression durch die Beschreibung des Zusammenhangs von Jargon, Aggression und Kultur*] (Goethe-Universität, Frankfurt, 2002); cf. también el más recinte *The Technological Republic: Hard Power, Soft Belief, and the Future of the West* (UK: Penguin, 2025), de Alexander C. Karp y Nicholas W. Zamiska.

[23] Cf. Pompeu Casanovas (2025). "La regimentación tecnológica: Inteligencia artificial, fascismo, agresión y sociedad democrática". *Teknokultura: Revista de Cultura Digital y Movimientos Sociales*, *22*(2), 137-149.

Esto tampoco es nuevo: William Graham Sumner (1840-1910) ya elaboró la teoría de la necesidad funcional de una élite de millonarios para el desarrollo social en "What the Social Classes Owe To Each Other" (1883). Lo importante es entender que Karp y Thiel no solamente son CEOs de grandes corporaciones. Son teóricos sociales y políticos. Hay en su obra un diseño social explícito y una teoría de la geopolítica global basada en la violencia necesaria para asegurar la supremacía de Estados Unidos como garante de los valores occidentales. Y la violencia también es un mercado.[24]

(iii) Quizás cabría añadir el cambio y el impacto en las actitudes agresivas y vindicatorias de una gran parte de la población en Internet, capaz de implementar, practicar y transmitir la violencia de la "cultura de la cancelación" como forma de sanción social.[25] Hay una misma aura moralista y excluyente en los tres tipos de reacción en el estado, el mercado y la población que llama la atención y debe ser interpretada. Este moralismo constituye el reverso de la ética que debería ser aplicada en los procesos de creación de valor social mediante la IA.

## 3. Segunda pregunta

**Ante los riesgos de la IA (sesgos, vigilancia masiva, manipulación), ¿qué ejemplos de regulaciones o políticas han demostrado que es posible fomentar innovación responsable, anteponiendo el interés público a la rentabilidad, sin ceder a la presión competitiva de actores como China o EE. UU?**

Podríamos completar la lista de riesgos mencionados por Albert Sabater con la lista de sesgos de una IA irresponsable ofrecida por Ricardo Baeza-Yates en su conferencia final de la Escuela de Verano: discriminación automatizada, pseudociencia, incompetencia humana, comercio digital injusto, desperdicio de recursos naturales y —especialmente después de la eclosión de la IA generativa con transformadores y LLM— violación de derechos de autor, desinformación y riesgos para la salud mental de las personas.[26] Muchos de los riesgos tienen relación con la humanización de los objetos a que somos propensos los humanos desde el nacimiento. La proyección de atributos humanos a las máquinas —*antropomorfismo*— especialmente (aunque no sólo) a robots y cobots, es una de las constantes de la crítica de Ramon López de Mántaras[27], de Carme Torras y Pablo Jiménez-Schlegel[28], y del mismo Baeza-Yates a la concepción simple de la IA. Las máquinas no "aprenden", no "alucinan", no "comprenden", no tienen "conciencia". Habría que añadir, también con Baeza-Yates, que la innovación no siempre es positiva. Resulta posible, asimismo —y de hecho así ocurre—, innovar en IA con fines bélicos o de otra índole que no están relacionados con la supuesta acribia del conocimiento científico. Hay que preguntar quién innova, cómo y para qué.

---

[24] Cuando consulté en septiembre del pasado año el valor de Palantir, éste alcanzaba los 180 billones (americanos) de dólares. En febrero, el valor de la acción había subido casi un cuarenta por ciento (76 $) y el valor de la compañía alcanzaba los 240 billones (USA). Después del nuevo panorama bélico en Ucrania e Israel, el valor de mercado de la compañía era aproximadamente de 370 billones y la acción se cotizaba a $ 156,14 USD (5 de septiembre de 2025). El 27 de septiembre, el valor de la compañía ha subido a 421,08 billones, y la acción a 177,57 $.
[25] He tenido ocasión de analizar el caso de Jon Tennant, por ejemplo, en Casanovas, P. (2025). "La perversidad inducida". *arXiv preprint arXiv:2503.23432*. https://arxiv.org/abs/2503.23432
[26] R. Baeza-Yates, "Las sombras de la inteligencia artificial y la responsabilidad de educar", AIHUB, Escuela de Verano, 4 de julio de 2025. Véase también, con el mismo título, "Las sombras de la IA" (conversación entre R. Baeza-Yates y Alec Dickinson, El Club de la IA, 19 de febrero de 2024), y "Las sombras de la IA y sus desafíos" (Congreso Futuro, 31 de enero 2024). Cf. el planteamiento teórico de Baeza-Yates en "Recomendaciones para una IA responsable", *Cuadernos de Beauchef* 8, no. 1 (2024): 99-111.
[27] Cf. Ramon López de Mántaras, "¿Es realmente inteligente la IA?" (Congreso Futuro, 31 de enero de 2024), *100 coses que cal saber sobre intel·ligència artificial,* Valls: Cossetània, 2023, trad. castellana, *100 cosas que hay que saber sobre intel·ligència artificial*, Barcelona: Lectio Ediciones, 2024.
[28] Cf. Carme Torras y Pablo Jiménez-Schlegl, *Robots. Una immersió ràpida*. Barcelona: Ed. Tibidabo, 2025.

Guardando esto en mente, la regulación puede ampararse en los instrumentos clásicos que ya hemos expuesto en la primera sección (*hard law*, *soft law*, políticas públicas, y ética), y desarrollar al mismo tiempo nuevos instrumentos para un uso responsable de la IA que vamos a denominar *gobernanza jurídica* de la IA.

Voy a poner dos tipos de ejemplos: (i) las denominadas "tecnologías cívicas" (*civic technologies*) de plataformas que se enfrentan a los problemas de las crisis políticas, desastres naturales y situaciones postbélicas (un ejemplo muy conocido es Ushaidi)[29]; (ii) los ecosistemas éticos y jurídicos que han empezado a crearse en algunos espacios de datos. Uno de los campos más prometedores es el de la denominada Industria 5.0, el espacio industrial de las manufacturas inteligentes, donde el ámbito de la implementación de estándares se está realizando desde la base, es decir, desde la creación de valor en el interior de cada manufactura. En cambio, en otros espacios de datos, por ejemplo, el financiero o el ámbito urbano, esto es más difícil de materializar, puesto que suele predominar el interés económico del banco o de la entidad financiera y, en el segundo, intervienen conflictos de política local.

Hemos denominado *legal over-compliance* (sobrecumplimiento jurídico) la estrategia de cumplir formalmente con el derecho, incluso más de lo exigido, para no poner en riesgo las inversiones o los beneficios y desviar la atención de obligaciones éticas o de gobernanza; por ejemplo, el servicio debido a todos los clientes o la adopción de las políticas de inclusión promovidas por el Banco Mundial.[30]

En mi opinión, deberíamos profundizar en el significado y en la operatividad de la denominada "cadena de valor" de la inteligencia artificial (*AI value chain*) en la creación de las condiciones para la emergencia de ecosistemas inteligentes éticos y jurídicos.[31] El Considerando n. 20 del Título Preliminar del Reglamento de IA (2024) apunta a la cadena de valor como la aportación fundamental de los sistemas de IA, y para ello, es esencial la educación o alfabetización en IA.[32] El Considerando n. 83 relaciona directamente la cadena de valor con los operadores en los niveles de riesgo.[33] Y ésta es también la perspectiva adoptada en el Código de Buenas Prácticas de la Ley de Inteligencia Artificial que ha entrado en vigor el 2 de agosto de 2025.

---

[29] Cf. https://www.ushahidi.com/
[30] Cf. de Koker, L., Casanovas, P. (2024). "'De-Risking', De-Banking and Denials of Bank Services: An Over-Compliance Dilemma?". In: Goldbarsht, D., de Koker, L. (eds) *Financial Crime, Law and Governance.* Ius Gentium: Comparative Perspectives on Law and Justice, vol 116. Springer, Cham, pp. 45-70. Open Access: https://doi.org/10.1007/978-3-031-59547-9_3
[31] Cf. Casanovas, P., Hashmi, M., De Koker, L., & Lam, H. P. (2025). "Compliance, Regtech, and smart legal ecosystems: a methodology for legal governance validation". In W. Barfield, U. Pagallo, *Research handbook on the law of artificial intelligence*. Cheltenmam, UK, Northhampton, USA: Edward Elgar Publishing, 2025, pp. 73–104. Un ecosistema jurídico puede definirse como un sistema complejo y dinámico que incluye múltiples niveles de gobernanza, desde el local hasta el nacional e internacional, y que involucra a una amplia gama de actores, incluidos legisladores, jueces, abogados, funcionarios encargados de hacer cumplir la ley, organizaciones de la sociedad civil y ciudadanos comunes. Un ecosistema jurídico inteligente funciona en un entorno que está (parcialmente) integrado en sistemas ciberfísicos y que abarca las características del Internet de las Cosas y de la Industria 4.0 y 5.0. (ética y derecho), lo que genera el cumplimiento en tiempo real.
[32] Considerando n. 20 AIA: "Con el fin de obtener los mayores beneficios de los sistemas de IA, protegiendo al mismo tiempo los derechos fundamentales, la salud y la seguridad, y de posibilitar el control democrático, la alfabetización en materia de IA *debe dotar a los proveedores, responsables del despliegue y personas afectadas de los conceptos necesarios para tomar decisiones con conocimiento de causa en*

Sin embargo, queda por determinar de qué forma puede comprobarse su efectividad, qué métricas pueden utilizarse, y cuál es la función de esta "cadena de valor" en el ciclo de vida de los sistemas de inteligencia artificial. La intuición es que ocupa un papel calibrador y equilibrador de regulación, puesto que es la utilidad y aceptación del sistema por parte de sus usuarios en los diversos contextos en los que opera lo que va a determinar su valor.

## 4. Tercera pregunta

**En un escenario donde los EE. UU priorizan la flexibilidad, ¿qué mecanismos podrían garantizar que la cooperación internacional en IA no se convierta en una carrera hacia el abismo en derechos, sino en un estándar global de responsabilidad?**

La respuesta a esta pregunta parece apuntar a los mecanismos de autoregulación, co-regulación y de control en la creación de valor: aquellos que promuevan la *proactividad* de los usuarios en sus tareas cotidianas como trabajadores, consumidores o ciudadanos; es decir, aquellos mecanismos que permitan la creación y el funcionamiento de la cadena de valor mediante una mejor coordinación de las actividades evitando la ejecución directa (*enforcement*). Por ejemplo: los *sandboxes* o espacios de prueba para la IA y la experimentación jurídica contemplados tanto por el Reglamento de IA (2024) como por las

---

*relación con los sistemas de IA.* Esos conceptos pueden variar en función del contexto pertinente e incluir el entendimiento de la correcta aplicación de los elementos técnicos durante la fase de desarrollo del sistema de IA, las medidas que deben aplicarse durante su uso, las formas adecuadas de interpretar los resultados de salida del sistema de IA y, en el caso de las personas afectadas, los conocimientos necesarios para comprender el modo en que las decisiones adoptadas con la ayuda de la IA tendrán repercusiones para ellas. En el contexto de la aplicación del presente Reglamento, *la alfabetización en materia de IA debe proporcionar a todos los agentes pertinentes de la cadena de valor de la IA los conocimientos necesarios para garantizar el cumplimiento adecuado y la correcta ejecución.* Además, la puesta en práctica general de *medidas de alfabetización en materia de IA y la introducción de acciones de seguimiento adecuadas podrían contribuir a mejorar las condiciones de trabajo y, en última instancia, sostener la consolidación y la senda de innovación de una IA fiable en la Unión.* El Consejo Europeo de Inteligencia Artificial (en lo sucesivo, «Consejo de IA») debe apoyar a la Comisión para promover las *herramientas de alfabetización en materia de IA*, la sensibilización pública y la comprensión de los beneficios, los riesgos, las salvaguardias, los derechos y las obligaciones en relación con el uso de sistemas de IA. En cooperación con las partes interesadas pertinentes, la Comisión y los Estados miembros deben facilitar la elaboración de códigos de conducta voluntarios para promover la alfabetización en materia de IA entre las personas que se ocupan del desarrollo, el manejo y el uso de la IA."

[33] Considerando n. 83 AIA: "Teniendo en cuenta la naturaleza y la complejidad de la cadena de valor de los sistemas de IA y de conformidad con el nuevo marco legislativo, es esencial garantizar la seguridad jurídica y facilitar el cumplimiento del presente Reglamento. Por ello es necesario aclarar la función y las obligaciones específicas de los operadores pertinentes de toda dicha cadena de valor, como los importadores y los distribuidores, que pueden contribuir al desarrollo de sistemas de IA. En determinadas situaciones, esos operadores pueden desempeñar más de una función al mismo tiempo y, por lo tanto, deben cumplir de forma acumulativa todas las obligaciones pertinentes asociadas a dichas funciones. Por ejemplo, un operador puede actuar como distribuidor e importador al mismo tiempo."

provisiones europeas para una mejor regulación (*Better Regulation*)[34], aunque éstos últimos parecen dirigirse más a la participación de los ciudadanos en la regulación que no a aspectos más técnicos de la construcción de la sociedad civil y el mercado digital europeo. Son ya comunes los estudios de impacto (*impact assessment*) y la apertura de rondas de opinión antes de la legislación. Pero lo que me parece más importante es la reconstrucción del espacio monetario, bancario y financiero emprendido por el reglamento de *Markets in Crypto-Assets Regulation* (MiCA).[35] Las cadenas de valor son un asunto central en MiCA, aunque no lo mencione directamente. Su estudio de impacto subraya el poder transformador de los activos virtuales (*cryptoassets*) y el papel potencialmente beneficioso de las cadenas de bloque en el sector financiero.[36]

El Reglamento abarca tres tipos de criptoactivos: (i) tokens referenciados a activos (ART, *asset-referenced tokens*)[37], (ii) tokens de dinero electrónico (EMT, *electronic money tokens*)[38], (iii) y otros criptoactivos no contemplados por la legislación vigente de la UE.[39] Está aún por ver su efectividad y eficacia; pero me parece importante apuntar el cambio que significa incorporar una autoridad descentralizada en los mercados. Desde el punto de vista de los estudios financieros y políticos, hay visiones muy críticas sobre lo que las cadenas de bloque y las criptomonedas significan para la economía.[40] Pero lo que

---

[34] EU Commission. *Better Regulations Guidelines*. Brussels, 3.11.2021, SWD (2021) 305 final. EU Commission. *Better Regulations Toolbox*. July 2023, complementing the better regulation guidelines presented in SWD(2021) 305 final.

[35] Regulation (EU) 2023/1114 of the European Parliament and of the Council of 31 May 2023 on markets in crypto-assets, and amending Regulations (EU) No 1093/2010 and (EU) No 1095/2010 and Directives 2013/36/EU and (EU) 2019/1937. Aplicable desde diciembre de 2024.

[36] Cf. Brussels, 24.9.2020 SWD(2020) 380 *Final Commission Staff Working Document Impact Assessment Accompanying the document Proposal for a Regulation of the European Parliament and of the Council on Markets in Crypto-assets and amending Directive (EU) 2019/1937*), p. 9. "Los criptoactivos y las DLT (*Digital Ledger Technologies*) subyacentes también ofrecen un gran potencial para mejorar la eficiencia en el sector financiero tradicional. Este potencial se deriva principalmente de dos características de la tecnología: (i) la capacidad de registrar información en un formato seguro e inmutable y (ii) la capacidad de hacer que esta información sea accesible de forma transparente para todos los participantes del mercado en la red DLT. La tokenización de valores (acciones o bonos) es un ejemplo de potencial de crecimiento a corto plazo. Esto puede conducir a una mayor financiación de las empresas mediante ofertas de tokens de valores (STO, *Security Tokens Offering*) *y a mejoras de eficiencia en toda la cadena de valor al reducir la necesidad de intermediarios y la automatización, lo que resulta en transacciones más rápidas, económicas y sin fricciones* [énfasis añadido]".

[37] Los *tokens referenciados a activos* (ART) son criptoactivos que buscan mantener un valor estable haciendo referencia a una canasta de activos, como monedas fiduciarias, materias primas u otros criptoactivos. Son comercializados a menudo como monedas estables.

[38] Los *tokens de dinero electrónico* son generalmente utilizados como medio de pago. Se trata de un tipo de criptoactivo diseñado para mantener un valor estable, generalmente vinculado a una única moneda fiduciaria como el euro o el dólar estadounidense.

[39] Por ejemplo, los NFTs (*Non-Fungible Token*) únicos y no intercambiables, las monedas digitales de bancos centrales (CBDC, *Central Bank Digital Currency*), y los criptoactivos considerados instrumentos financieros tradicionales regulados por MiFID II (*Markets in Financial Instruments Directive* II). La normativa de revisión de MiFID II/MiFIR entró en vigor el 28 de marzo de 2024, mientras que la fecha límite de transposición a los estados nacionales de las modificaciones de MiFID II es el l 29 de septiembre de 2025. Cf. DIRECTIVA 2014/65/UE DEL PARLAMENTO EUROPEO Y DEL CONSEJO de 15 de mayo de 2014 relativa a los mercados de instrumentos financieros y por la que se modifican la Directiva 2002/92/CE y la Directiva 2011/61/UE (refundición). (Texto pertinente a efectos del EEE).(DO L 173 de 12.6.2014, p. 349)

[40] Cf Fonghu Z, Yuen TH. "A critique on decentralized finance from a social, political, and economic perspective". *Blockchain* 2023(1):0006, https://doi.org/10.55092/blockchain20230006. Los autores sintetizan los riesgos que han sido repetidamente señalados por la crítica: (1) una obsesión equivocada con las nociones libertarias, (2) incentivos insostenibles promulgados y reforzados por mecanismos DeFi que crean y

aparece ya de forma clara es que es urgente la implementación de una regulación consensuada que no niegue la realidad de un mercado que mantiene un crecimiento continuo. Especialmente después del desmantelamiento de las administraciones, el estado y la centralización de la Reserva Federal "bajo tutela" impulsados por la agresiva política económica de Trump.[41] El equilibrio entre centralización y descentralización no es equivalente al de regulación y desregulación que había sido hasta ahora el marco general de la discusión. Hay que reconocer y debatir este nuevo marco, que comprende también una nueva tensión por la soberanía digital entre Europa y Estados Unidos.[42]

Es esta perspectiva sobre la regulación lo que merece la pena tener en cuenta para la educación ética de los maestros, investigadores y ciudadanos. Para que la IA estabilice un estándar global de responsabilidad no se trata de aplicar ahora políticas internacionales desde arriba en una defensa a ultranza del estado de derecho mediante leyes y políticas públicas ejecutables. Esto debe darse por supuesto, y con estos instrumentos solamente no creo que pueda garantizarse una situación regulatoria sostenible en el panorama que he descrito anteriormente. No se trata ahora de moralizar, sino de explicar y ofrecer razones para que investigadores y empresas puedan emprender diseños innovadores con confianza. Es decir, debería empezarse desde abajo y desde el núcleo de conceptos que se hallan en el centro de la construcción de las plataformas (*middleware*): una perspectiva desde y para los ciudadanos, empresas y consumidores a nivel micro y meso, i.e., desde el interior de los requerimientos que deben respetar los procesos de información en las plataformas de tratamiento de datos en los ámbitos de los espacios, tanto privados como comunes, que constituyen la denominada "economía de plataforma" (*platform-driven economy*).

---

comercializan *borderline* esquemas Ponzi Web 3, y (3) una tendencia a la centralización y la extorsión de activos socialmente aceptable en momentos de tensión. Así las criptomonedas ya no serían solamente un método de pago que protege la privacidad, sino vehículos para la especulación incesante e influencias extremadamente altos, y los NFT ya no son simplemente activos digitales únicos, sino cuasivalores ofrecidos principalmente por individuos que pueden, en cualquier momento y sin consecuencias, huir con los fondos. Véase asimismo, Kelsie Nabben (2023) "Cryptoeconomics as governance: An intellectual history from 'Crypto Anarchy' to 'Cryptoeconomics', *Internet Histories*, 7:3, 254-276, DOI: 10.1080/24701475.2023.2183643.

[41] Véanse las intervenciones (entre otros la del propio Jerome Powell) en el reciente *Jackson Hole Economic Policy Symposium: Labor Markets in Transition — Demographics, Productivity and Macroeconomic Policy* (August 21, 2025 – August 23, 2025). https://www.kansascityfed.org/research/jackson-hole-economic-symposium/2025/

[42] F. Bria, P. Timmers, F. Gernone (2025). *EuroStack – A European Alternative for Digital Sovereignty*, Bertelsmann Stiftung, Gütersloh February 2025, pp- 50-51. "Iniciativas como los *Projects of Common European Interest in Cloud Infra-structure and Services* (PCEI-CIS) están allanando el camino para infraestructuras descentralizadas y federadas capaces de procesar datos más cerca de su origen. Este enfoque mejora la resiliencia, reduce la latencia y refuerza la seguridad. La computación de borde (*Edge-computing, procesamiento local de datos*) ha trascendido las discusiones teóricas y ahora se está implementando activamente en sectores como la conducción autónoma, las fábricas inteligentes y la atención médica, lo que sitúa a Europa a la vanguardia de estas tecnologías transformadoras. Sin embargo, la tecnología *blockchain*, que sustenta gran parte del movimiento actual de descentralización y el llamado paradigma de la web 3.0, sigue estando dominada por actores estadounidenses. Europa aún no ha recuperado su posición en este ámbito crítico, lo que supone un desafío para sus ambiciones de descentralización más amplias."

## 5. Implicaciones para la educación de maestros, profesores e investigadores

La educación de maestros e investigadores podría realizarse en la dirección sugerida por la respuesta a las preguntas anteriores. Es crucial un conocimiento más preciso de las condiciones en que se produce la elaboración y entrenamiento de programas de IA generativa y la situación geopolítica estratégica global. Deberíamos reflexionar sobre un conjunto de estrategias que cubran todo el espectro de lo que resulta relevante en los distintos escenarios y niveles educativos, i.e. en el aula, en la interacción con los estudiantes, en la preparación de los marcos educativos de instrucción y prácticas, en la propia preparación y visión de los educadores. Hay que tomar algunas precauciones también contra la ingenuidad ético-política, y el posible desconocimiento de los orígenes, contingencia y efectos de la economía de plataforma en los usuarios. Conocer, en definitiva, el marco de actuación es ya una condición imperativa para maestros, profesores e investigadores. La proactividad necesaria en el uso de instrumentos de inteligencia artificial depende de ello.

Hay en estos momentos una investigación ingente en IA y Educación (*AI and Education*, AIED). Existen múltiples encuestas bibliográficas[43], tipologías[44], análisis y metaanálisis en este campo.[45] La mayoría de los artículos efectúa una valoración positiva de su uso y señala tanto sus ventajas como sus limitaciones. Casi todos insisten en la dimensión ética como necesaria y muestran de qué modo está ya siendo considerada en los planes de estudio.

---

[43] Cf. Che Ghazali, Roshasfarizan, Mohd Fadzil Abdul Hanid, Mohd Nihra Haruzuan Mohd Said, and Huan Yik Lee (2025). "The advancement of Artificial Intelligence in Education: Insights from a 1976–2024 bibliometric analysis." *Journal of Research on Technology in Education* (2025): 1-17.; Holmes, Wayne, and Ilkka Tuomi (2022). "State of the art and practice in AI in education." *European journal of education* 57 (4): 542-570.

[44] Por ejemplo, Holmes y Tuomi (2022, 550) proponen la siguiente taxonomía: "(1) *IAED CENTRADA EN EL ALUMNO*: (i) Sistemas Inteligentes de Tutoría (ITS), (ii) Aplicaciones Asistidas por IA (p. ej., matemáticas, texto de voz, aprendizaje de idiomas), (iii) Simulaciones Asistidas por IA (p. ej., aprendizaje basado en juegos, VR, AR), (iv) IA para Apoyar a Estudiantes con Discapacidades, Redacción Automática de Ensayos (AEW), (v) Chatbots, (vi) Evaluación Formativa Automática (AFA), (vii) Orquestadores de Redes de Aprendizaje, (viii) Sistemas de Tutoría Basados en Diálogos (DBTS), (ix) Entornos de Aprendizaje Exploratorio (ELE), (x) Asistente de Aprendizaje Permanente Asistido por IA; (xi) *IAED CENTRADA EN EL PROFESOR*: (i) Detección de plagio, (ii) Cuidado Inteligente de Materiales de Aprendizaje, (iii) Monitoreo del Aula, (iv) Evaluación Sumativa Automática; (v) Asistente de Enseñanza con IA (incluido el asistente de evaluación), (vi) Orquestación del Aula; 3. *IAED DIRIGIDA A LA INSTITUCIÓN*: (i) Admisiones (p. ej., selección de estudiantes), (ii) Planificación de cursos, programación, horarios, (iii) Seguridad Escolar, (iv) Identificación de fracasos y estudiantes en riesgo, (v) Supervisión electrónica.

[45] Cf. Chen, Lijia, Pingping Chen, and Zhijian Lin (2020). "Artificial intelligence in education: A review." IEEE access 8: 75264-75278; Huang, Jiahui, Salmiza Saleh, and Yufei Liu. "A review on artificial intelligence in education." *Academic Journal of Interdisciplinary Studies* 10, no. 3 (2021): 206-2017; Yan, Lixiang, Lele Sha, Linxuan Zhao, Yuheng Li, Roberto Martinez-Maldonado, Guanliang Chen, Xinyu Li, Yueqiao Jin, and Dragan Gašević (2024). "Practical and ethical challenges of large language models in education: A systematic scoping review." *British Journal of Educational Technology* 55 (1): 90-112; Yu, Ji Hyun, Devraj Chauhan, Rubaiyat Asif Iqbal, and Eugene Yeoh (2025). "Mapping academic perspectives on AI in education: Trends, challenges, and sentiments in educational research (2018–2024)." *Educational technology research and development* 73 (1): 199-227; Kerimbayev, Nurassyl, Karlygash Adamova, Rustam Shadiev, and Zehra Altinay (2025) "Intelligent educational technologies in individual learning: a systematic literature review." *Smart Learning Environments* 12 (1): 1. Zhendong Chu, Shen Wang, Jian Xie, Tinghui Zhu, Yibo Yan, Jinheng Ye4, Aoxiao Zhong, Xuming Hu, Jing Liang, Philip S. Yu, Qingsong Wen1 (2025). "LLM Agents for Education: Advances and Applications" 2018. ACM ISBN 978-1-4503-XXXX-X/2018/06.

No faltan tampoco voces críticas que advocan por la supresión o la limitación severa de la IA en la práctica educativa. Así, un grupo de investigadores ha publicado recientemente (junio de 2025) una carta abierta para limitar la penetración y uso indiscriminado de la IA en la investigación y en las universidades que ya ha recibido 1339 adhesiones (a 27 de septiembre 2025).[46] En el polo contrario, se hallan los que defienden la necesidad de no ignorar el hecho de que la calidad instrumental de la IA debe ser reconocida también por su valor educativo y resulta ya imprescindible y urgente su plena integración en los *syllabi* y *curricula* de alumnos y docentes.[47] El reciente Informe ODITE (2025) sobre la función de las técnicas de IA en el aprendizaje personalizado, resalta este punto y subraya que

> no basta con una educación que nos dote de conocimientos, capacidades y habilidades (*cualificación*), o que nos socialice en las tradiciones, culturas, formas de ser y actuar (*socialización*). Necesitamos también, o más, si cabe, una educación que nos permita convertirnos en sujetos autónomos e independientes, agentes de nuestra propia vida (*subjetivación*).[48]

Aprendizaje *individualizado* y aprendizaje *personalizado* constituyen conceptos distintos. El primero refiere a un enfoque de la instrucción que adapta las experiencias educativas a las necesidades, objetivos y habilidades únicas de cada alumno. Pero personalizar el aprendizaje "implica además identificar los conocimientos previos de las personas que aprenden y fomentar la autoconciencia sobre sus necesidades individuales de aprendizaje".[49] Hay otros conceptos —como el 'generación aumentada por recuperación'[50] — que emergen para materializar esta perspectiva. Y esto desde la educación primaria.

---

[46] Open Letter: Stop the Uncritical Adoption of AI Technologies in Academia . Cf. Guest, Olivia, Marcela Suarez, Barbara Müller, Edwin van Meerkerk, Arnoud Oude Groote Beverborg, Ronald de Haan, Andrea Reyes Elizondo et al. (2025). Against the Uncritical Adoption of'AI Technologies in Academia. "La industria tecnológica se aprovecha de nosotros, a veces incluso hablando a través nuestro, para convencer a nuestros estudiantes de que estas tecnologías de IA son útiles (o necesarias) y no perjudiciales. Por lo tanto, argumentamos que los líderes y administradores universitarios deben actuar para ayudarnos colectivamente a frenar la ola de software basura, que alimenta clichés dañinos (por ejemplo, los llamados estudiantes perezosos) y marcos falsos (por ejemplo, la supuesta eficiencia o inevitabilidad) para obtener penetración en el mercado y aumentar la dependencia tecnológica." Agradezco a Txetxu Ausín el conocimiento de esta iniciativa.
[47] Cf. Ray Gallon, "Incorporar la IA en un entorno de formación profesional superior", en el Informe ODITE, J.M. Muñoz, N. Lorenzo, X. Suñé, M.A. Prats (coords.) (2025). *Inteligencias conectadas: cómo la IA está redefiniendo el Aprendizaje Personalizado*. Almería: ODITE, Espiral, pp. 260-269. Galon escribe: "El autor de este estudio considera que, si no se incorpora la intel·ligència artificial y su uso docente en nuestro plan de estudiós y en los procesos de gestión de contenido académico, no estaremos cumpliendocon nuestro compromiso como educadores, y deberíamos avergonzarnos de nuestra negligència" (ibid. p. 262).
[48] Cf. Carlos Magro Mazo, "Sí. Esta vez sí funcionará", en en el Informe ODITE, J.M. Muñoz, N. Lorenzo, X. Suñé, M.A. Prats (coords.) (2025). *Inteligencias conectadas: cómo la IA está redefiniendo el Aprendizaje Personalizado.* Almería: ODITE, pp. 24-37 (ibid. p. 31).
[49] Ingrid Noguera, Anna Ciraso-Cali, Lucía Catarineu (2025). "Personalización del aprendizaje: perspectivas del alumnado y el profesorado sobre los usos de la IA en la Universidad", en el Informe ODITE, J.M. Muñoz, N. Lorenzo, X. Suñé, M.A. Prats (coords.) (2025). *Inteligencias conectadas: cómo la IA está redefiniendo el Aprendizaje Personalizado.* Almería: ODITE, pp. 192-205 (ibid. p. 193).
[50] Cf. Por ejemplo, la 'Generación Aumentada por Recuperación' (*Retrieval Augmented Generation* RAG), refiere a un enfoque controlado mediante el uso de IA que permite la generación de información aumentada por la recuperación de contenidos seleccionados en fuentes protegidas Neus Lorenzo (2025). "Las RAG, o el nacimiento de los ecosistemas IA personalizados para una intervención educativa de calidad: el papel de ODITE", en en el Informe ODITE, J.M. Muñoz, N. Lorenzo, X. Suñé, M.A. Prats (coords.) (2025).

Respecto a la dimensión regulatoria, podemos identificar tres puntos que resultan relevantes y se repiten en la literatura existente: (i) la insistencia en la importancia de la dimensión ética; (ii) la crítica a la débil preocupación de las políticas educativas de los estados nacionales por los efectos de la inteligencia artificial como método de aprendizaje (lo que explicaría la reacción de los investigadores que acabamos de citar); (iii) la diversidad de marcos legales para hacer frente al impacto de chatbots, LLM y ChatGPT en las instituciones educativas. Esto último constituye un serio inconveniente, puesto que no existe ni una estrategia unificada de regulación, ni un consenso general sobre cómo sería ello posible.

Pero quizás podríamos aprovechar también este hecho para insistir en la capacidad de los propios estudiantes, profesionales y administraciones de investigación y educación para proponer marcos que no tienen por qué ser generales, sino más concretos, locales y regionales. Esto tiene la ventaja de partir de la experiencia real de las aulas, laboratorios y centros de investigación. Creemos que no es solamente mediante la participación sino mediante la proactividad, resiliencia e imaginación de los usuarios, diseñadores y creadores que puede regularse de forma efectiva el uso académico (en sentido amplio) de la IA.

Hay ya algunas propuestas específicas que van en este sentido. De acuerdo con el reciente *Manifiesto en defensa de una educación centrada en humanos en la era de la IA* (2025), resulta posible distribuir el uso creativo de la IA en distintos niveles de compromiso (Tabla 1). Esto va a resultarnos útil en la próxima sección.

**Tabla 1.** Seis niveles de participación creativa en la interacción humano/a-IA en educación. Fuente: Romero et al. (2025).[51]

| Niveles | Grado de compromiso del alumno con el uso de la IA |
|---|---|
| 1. Consumidor pasivo | El alumno consume contenido generado por IA sin comprender su funcionamiento. |
| 2. Consumidor interactivo | El alumno interactúa con el contenido generado por IA. El sistema de IA se adapta a sus acciones. |
| 3. Creación individual de contenido | El alumno crea contenido nuevo utilizando herramientas de IA. |
| 4. Creación colaborativa de contenido | Un equipo crea contenido nuevo utilizando herramientas de IA. |
| 5. Co-creación participativa de conocimiento | Un equipo crea contenido gracias a herramientas de IA y a la colaboración de las partes interesadas en un problema complejo. |
| 6. Aprendizaje expansivo con apoyo de la IA | En intervenciones formativas con apoyo de IA, la acción de los participantes puede ampliar o transformar situaciones problemáticas. |

---

*Inteligencias conectadas: cómo la IA está redefiniendo el Aprendizaje Personalizado.* Almería: ODITE, pp. 50-59.

[51] Margarida Romero, Thomas Frosig, Amanda M. L. Taylor-Beswick, Jari Laru, Bastienne Bernasco, Alex Urmeneta, Oksana Strutynska, and Marc-André Girard (2025). "Manifesto in Defence of Human-Centred Education in the Age of Artificial Intelligence", p. 158-170.

Una explicación viable de la tabla 1 sería la siguiente, propuesta por la misma autora:

> La Tabla 1 permite visualizar diferentes niveles de usos de la IA en educación, que transitan desde formas de dependencia tecnológica hacia escenarios de co-creación y transformación colectiva. Este tránsito, sin embargo, no ocurre de manera automática: exige competencias interpretativas, éticas y metacognitivas que solo pueden adquirirse mediante una alfabetización en IA (*AI literacy*) orientada al pensamiento crítico, la comprensión de los mecanismos de los modelos generativos y el uso responsable de la IA. En esta línea, la alfabetización en inteligencia artificial constituye un requisito indispensable para preservar y ampliar la *agentividad* de docentes y discentes, entendida como la capacidad de actuar de manera informada, crítica y creativa dentro de entornos digitales cada vez más mediados por sistemas algorítmicos. Por ello, la formación en IA no debería limitarse al ámbito escolar, sino expandirse hacia la educación postsecundaria, la formación profesional y el aprendizaje a lo largo de la vida. La alfabetización en IA se configura como el nuevo fundamento de una ciudadanía pedagógica capaz de mantener la agencia humana dentro de los ecosistemas de conocimiento actuales.[52]

Volvamos ahora a las preguntas plateadas por Albert Sabater.

## 6. Vuelta a las preguntas iniciales.

### 6.1 ¿Cómo pueden los marcos regulatorios priorizar la protección de derechos fundamentales en el desarrollo de IA, sin caer en la falsa dicotomía entre regulación e innovación?

Lo primero es conocerlos bien y reconocer tanto su potencial como sus limitaciones desde una visión crítica de la legislación aplicable. Quizás deberíamos subrayar la complejidad de los marcos regulatorios actuales, donde los instrumentos jurídicos clásicos (legislación y jurisprudencia, principalmente) se apoyan en los nuevos instrumentos de "derecho flexible" (*soft law*) y en el consenso de múltiples actores para resultar operativos. Así, la UE está desarrollando actualmente estándares para el uso de la IA en todos los ámbitos, del mismo modo que las grandes asociaciones profesionales han desarrollado globalmente los estándares tecnológicos para la gobernanza de corporaciones, industrias y manufacturas.

Esto parece una clara indicación a favor del comportamiento participativo de todos los actores (ciudadanos, consumidores, empresas…), en la línea de la "nueva legislación" europea. Pero tiene costes también, puesto que los que resultan más favorecidos son los que ya lo son en el mercado. Así, en su análisis de la dimensión externa de los estándares de inteligencia artificial en la política legislativa europea, H.W. Micklitz ha sugerido la alegoría de la liebre y la tortuga para describir la situación. Mientras la industria global

---

[52] Margarida Romero, comunicación personal. Agradezco a la autora la aportación de esta explicación. A ello cabría añadir en el punto 6 el comentario de Neus Lorenzo relativo al impacto retroactivo de la IA en el propio equipo educativo, puesto que la creación de vínculos con agentes de IA afecta también a la construcción de la experiencia emocional entre humanos

ha ido desarrollando con el tiempo unos estándares muy detallados para cada sector, más centrados en los aspectos técnicos que en los derechos humanos, la UE está intentado emularlos con cierta prisa, descuidando también la dimensión necesariamente ejecutiva (*enforceable*) que la protección e implementación de los derechos humanos requiere.[53]

Es más fácil imponerlos cuando se convierten en derechos fundamentales, i.e. derechos insertos en los textos constitucionales nacionales. No siempre es así, y ello provoca problemas para su materialización, porque quedan entonces al amparo del derecho internacional público. Esto sucede no solamente en Arabia Saudita, Pakistán, e Irán, sino en países de la *Commonwealth* como Australia, con frecuentes críticas de la doctrina jurídica interna sobre las consecuencias que tiene, por ejemplo, en materia de inmigración y de protección de los derechos de los refugiados.[54]

Respecto al uso de la IA, la propuesta sería reconocer la necesidad de combinar y precisar las dos dimensiones o ejes —vertical y horizontal (Fig. 1)— del estado de derecho en una gobernanza jurídica que tenga en cuenta tanto la proactividad y capacidad auto y co-reguladora de los usuarios (no solamente su 'participación') como el hecho de que la materialización de los derechos humanos no es fácilmente adoptada por los estados mediante normas con valor ejecutivo. Los derechos deben poder ser defendidos activamente por los usuarios. Y por 'usuario' aquí entendemos tanto ciudadanos como empresas y administraciones. Es necesaria una política legislativa más decidida en este sentido, y las posiciones no son unitarias dentro del mercado ni en el espacio político. Hay posiciones distintas en el interior de los estados, empresas y corporaciones.

### 6.2 Ante los riesgos de la IA (sesgos, vigilancia masiva, manipulación), ¿qué ejemplos de regulaciones o políticas han demostrado que es posible fomentar innovación responsable, anteponiendo el interés público a la rentabilidad, sin ceder a la presión competitiva de actores como China o EE. UU?

De acuerdo con los estudios bibliométricos citados anteriormente, Estados Unidos y China son precisamente los dos países donde más se ha desarrollado académicamente el campo de educación y IA (IAED). En la revisión sistemática reciente de la literatura sobre sistemas multiagente y educación, se ha distinguido entre: (i) agentes *pedagógicos*, que se centran en la automatización de tareas complejas para apoyar tanto a docentes como a estudiantes; y (2) agentes *educativos de dominio específico*, diseñados para campos especializados como la educación científica, el aprendizaje de idiomas y el desarrollo profesional.[55] En ambos casos se comprueba que los retos radican en "la falta de marcos estructurados para integrar a los agentes de LLM en los sistemas educativos", de manera que puedan funcionar de manera homogénea y regulada.[56] "Las investigaciones futuras

---

[53] Micklitz, H. W. (2024). The External Dimension of AI Standards in EU Digital Policy Legislation. *Texas International Law Journal*, *60* (1): 129-149.
[54] Cf. e.g. Patrick Keyzer (2025). *Impunity, Meet Dignity. Listening, Justice, Open Rights, and A.I. in Australia*. Cham: Springer, en prensa.
[55] Chu, Zhendong, Shen Wang, Jian Xie, Tinghui Zhu, Yibo Yan, Jinheng Ye, Aoxiao Zhong et al. (2025) "LLM agents for education: Advances and applications." arXiv preprint arXiv:2503.11733.
[56] "Si bien modelos como el marco FOKE (*Forest Of Knowledge and Education framework*) combinan modelos básicos, grafos de conocimiento e ingeniería de indicaciones (*prompts*) para ofrecer servicios de aprendizaje interactivos y explicables, una adopción más amplia requiere modelos escalables que puedan validarse en diversos entornos educativos del mundo real. Además, los LLM se han explorado como

deberían centrarse en el desarrollo de marcos estandarizados para guiar la implementación estructurada de los LLM en el aprendizaje personalizado" (*ibid*.).

En la línea de argumentación seguida hasta aquí, hemos subrayado el hecho de que las regulaciones y estrategias compartidas pueden ofrecer resultados transversales que la opción vertical de la regulación no está en condiciones de ofrecer. Esta colaboración se da en el meso-nivel de los conceptos aplicados y las micro-situaciones y escenarios donde es posible identificar a los actores y agentes para lograr un resultado común.

La educación constituye un ámbito que puede ilustrar este aspecto. El *Informe ODITE* y el *Manifiesto* citados en la sección anterior ofrecen muchos ejemplos de actuaciones, programas y aplicaciones creados y co-creados en el aula para que alumnos y profesores constituyan una única unidad dinámica o dialógica de aprendizaje. Se trata de una aproximación endógena, interna a la experiencia y al proceso educativo.

Pero también puede adoptarse una aproximación exógena, i.e. externa y complementaria, donde la innovación radica en la metodología utilizada para lograr resultados fiables y enriquecer el conocimiento que ahora se tiene de problemas como el del abandono o el fracaso escolar. La imaginación en formular estrategias de uso de la IA para mejorar el conocimiento puede llevar a formas innovadoras de comprensión que conduzcan a nuevos escenarios de diálogo. Por ejemplo, la tutela y monitorización individual de los estudiantes en esta situación pueden verse beneficiados por el resultado de técnicas de LLM y ML. Así, mediante la explotación de una base de datos que contiene 124.00 registros de estudiantes según 36 características y una metodología de comparación de subpoblaciones (estudiantes, y estudiantes bajo tutela), investigadores del EC de Monterrey ha descubierto que las explicaciones sobre las razones del abandono estudiantil obtenidas coincidían con las experiencias reales de mentores y tutores, proporcionando nuevos puentes de diálogo.[57] Más aún, puede implementarse el aprendizaje automático cuántico (QML) en datos educativos para predecir, por ejemplo, los resultados y la proyección posterior de los exalumnos.[58]

Como indican los mismos autores, no ha habido una penetración real aún de este tipo de técnicas en la educación. Mi argumento aquí, para responder a la segunda pregunta de Albert Sabater, es que la innovación responsable y el interés público pueden conyugarse a partir de la actuación de los investigadores, profesores y alumnos implicados en los mismos centros donde ocurren los avances y los abandonos. La distinción radical entre el

---

herramientas para potenciar la creatividad y la colaboración en el aprendizaje basado en problemas, apoyando a los estudiantes mediante la lluvia de ideas, la resolución de problemas y la ejecución de proyectos. *Sin embargo, los estudios indican que su eficacia se ve limitada por la ausencia de marcos de orientación estructurados que ayuden a educadores y estudiantes a incorporar sin problemas los agentes de LLM en los flujos de trabajo del PBL (Project-based learning)*." (*ibid*., énfasis nuestro) Cf. asimismo Hu, S. and Wang, X., 2024, August. [Foke: A personalized and explainable education framework integrating foundation models, knowledge graphs, and prompt engineering](#). In *China National Conference on Big Data and Social Computing*. Singapore: Springer Nature Singapore, pp. 399-411.

[57] Talamás-Carvajal, Juan Andrés, Héctor G. Ceballos, and Isabel Hilliger (2025). "[The Facts Behind the Prophecy: Validating a Methodology for Identifying Behavioural Differences in Higher Education Student Subpopulations Under Intervention](#)." *Journal of Learning Analytics* 12 (2): 1-13.

[58] Ramos-Pulido, Sofía, Neil Hernández-Gress, Glen S. Uehara, Andreas Spanias, and Héctor G. Ceballos-Cancino (2015). "Implementation of Quantum Machine Learning on Educational Data." In *International Conference on Agents and Artificial Intelligence (ICAART)*, vol. 3, pp. 480-487. Science and Technology Publications, Lda, 2025.

interés público y el privado tiende a desvanecerse en contextos como éstos, en los que la interpenetración entre ambos es una manifestación más de la sociedad híbrida (entre humanos/as y máquinas) en la que ya nos encontramos. Son los propios sujetos implicados los que tienden a reaccionar positivamente ante los problemas que realmente tienen.

### 6.3 En un escenario donde los EE. UU priorizan la flexibilidad, ¿qué mecanismos podrían garantizar que la cooperación internacional en IA no se convierta en una carrera hacia el abismo en derechos, sino en un estándar global de responsabilidad?

Una visión crítica de la legislación aplicable y de la propia función educativa puede ayudar a tener una mejor comprensión del trabajo por hacer. Como mínimo, sin integrar por el momento el marco legislativo en China, hay tres tipos de regulaciones cuyo tratamiento de la IA depende de la distinta cultura jurídica en que se dan. Hay por lo menos tres estrategias distintas, que corresponden al distinto tipo de cultura jurídica de Europa, Estados Unidos y los países de la Commonwealth: (i) La UE opta por un *enfoque integral*, mediante la mencionada Ley de IA, que ha creado un marco jurídico único aplicable a todos los sectores; (ii) UK ha diseñado una estrategia "pro-innovación" basada en *principios* —seguridad, transparencia, equidad, rendición de cuentas y refutabilidad)— para que los reguladores los interpreten y apliquen en cada sector específico[59]; (iii) EEUU se basa en un modelo *sectorial*, un «mosaico de regulaciones» (*patchwork kilt*) de leyes ya en vigor, estándares NIST, órdenes presidenciales y órdenes ejecutivas sectoriales.

En el caso de UK y EEUU, hay una cierta dependencia del partido de gobierno. Europa, en cambio, mantiene una perspectiva vertical centralizada propiciada por la opción de regular la IA mediante directivas y reglamentos. Pero esto también presenta muchas dificultades. La revisión del Informe Draghi elaborada y presentada por el mismo Draghi en la Conferencia de revisión de su informe sobre la competitividad de 16 de septiembre de 2025, aboga explícitamente por una simplificación del reglamento de privacidad (GDPR) y una adaptación de los estrictos requisitos impuestos a los diseñadores, proveedores y administradores de sistemas de IA para un desarrollo más ordenado y competitivo del mercado digital europeo.[60] Resulta significativo que subraye la creación y la cadena de valor por parte de la IA.[61] Los sistemas deben integrarse en el tejido productivo, el sistema financiero y los espacios de datos con su potencialidad propia de mejora.

En lo que atañe a la tercera pregunta, creo que una posible respuesta en línea con las anteriores es insistir en la capacidad creativa de los actores del campo de la educación, actuando desde la base, sin esperar una regulación *ad hoc*. La sugerencia es aquí aprovechar las experiencias y modelos existentes provenientes de los propios profesionales e investigadores especializados que puedan contribuir a elaborar, visualizar e implementar la cadena de valor de la IA en el ámbito educativo. Pienso en el modelo gradual de la Fig.1, que podría ser considerado para una regulación trasversal y efectiva desde la base. Pero elaborar seriamente este argumento cae ya fuera del objeto y propósito del presente artículo. Baste por ahora indicar que una regulación responsable no puede darse fuera de los mecanismos de gestión, control y experimentación de aquellos a los que va dirigida.

---

[59] Cf. A pro-innovation approach to AI regulation. Updated 3 August 2023 (published under the 2022 to 2024 Sunak Conservative government).

[60] Mario Draghi (2025). European Commission. *High Level Conference – One year after the Draghi report: What has been achieved, what has changed,* September 16th 2025.

[61] "En los próximos meses, el sector de la automoción pondrá a prueba la capacidad de Europa para alinear la regulación, la infraestructura y el desarrollo de la cadena de suministro en una estrategia coherente para una industria que emplea a más de 13 millones de personas en toda la cadena de valor". Draghi (2025: p. 7).

## Agradecimientos

Pompeu Casanovas

IIIA-CSIC, Bellaterra, Barcelona, 7 de octubre de 2025.